\documentclass{article}
\usepackage{isita2008}
\usepackage[english]{babel} 
\usepackage{algorithm}
\usepackage{algorithmic}
\newcommand{\B}[1]{\boldsymbol{#1}}

\usepackage{amsmath}
\usepackage{amssymb}
\usepackage{epsfig}
\usepackage{psfrag}
\DeclareMathOperator*{\argmin}{argmin}
\DeclareMathOperator*{\argmax}{argmax}

\begin{document}
\selectlanguage{english}

\title{Achieving near-Capacity on Large Discrete Memoryless Channels\\
 with Uniform Distributed Selected Input}
\name{Amine Mezghani, Michel T. Ivrla\v{c} and Josef A. Nossek%
\thanks{This work
was supported by the deutsche Forschungsgemeinschaft (DFG) under Grant SPP1202.}
}

\address{
\begin{tabular}{c}
Institute for Circuit Theory and Signal Processing\\
Technische Universit\"at M\"unchen\\
Theresienstra\ss e 90,  80290 Munich, Germany\\
E-mail: \{Mezghani, Ivrlac, Nossek\}@nws.ei.tum.de
\end{tabular}
}
\sloppy
\maketitle 
\begin{abstract}
We propose a method to increase the capacity achieved by uniform prior in discrete memoryless channels (DMC) with high input cardinality. It consists in appropriately  reducing the input set. Different design criteria of the input subset are discussed. We develop an efficient algorithm to solve this problem based on the maximization of the cut-off rate. The method is applied to a mono-bit transceiver MIMO system,  and it is shown that the capacity can be approached within tenths of a dB by employing standard binary codes while avoiding the use of distribution shapers.
\end{abstract}
\section{INTRODUCTION}
The challenge with nonsymmetric and/or nonbinary channels is that the capacity-achieving probability distribution is not uniform \cite{mceliece}. In this case, distribution shapers are needed to approach capacity, which results in very large block sizes \cite{bennatan}. This is of course impractical for channels with low complexity receiver or where the sender and receiver wish to communicate without substantial average delay. To avoid distribution shapers, we require that all signals are used evenly.  In  \cite{Shulman}, it is shown that the degradation
in using uniform prior, instead of the capacity achieving distribution, is worst for the
Z-channel, and the amount of the degradation, for binary-input channels,
is quite small. For general DMCs, and especially those with high input cardinality,  we show in this paper that  uniform capacity can be increased by using a reduced packing of symbols. 
To this end, we try to find the best subset $\mathcal{X}'$ from the original input  set $\mathcal{X}$ so that all its symbols are distinguishable at the receiver and maximally spaced. This may be a crucial approach especially in large DMC channels, where the transmitter have somehow an access to the channel state information, by means of a feedback channel for example (or the channel is a priori known).
We could also require the subset size $|\mathcal{X}'|=K$ to be a power of 2 if a binary code is employed so that encoded bits can be directly mapped to the channel input symbols.\\
Our paper is organized as follows. First we formulate the problem mathematically based on different criteria in Section~\ref{section:scmodel}. Then we solve the problem of the optimal subset search based on two different criteria, respectively in Section~\ref{ser} and~\ref{section:cutoff}. In Section~\ref{application} we apply our method to a mono-bit  multi-input multi-output (MIMO) channel and show its  usefulness. Finally, we test the performance of the selected input subset when combined with an LDPC code under this kind of channels in Section~\ref{ldpc}.
\label{section:introduction}
\section{SYSTEM MODEL AND PROBLEM FORMULATION}
\label{section:scmodel}
 We consider a DMC with finite input alphabet $\mathcal{X}$ having the cardinality $|\mathcal{X}|=M$ and finite output
alphabet $\mathcal{Y}$. We assume the input to the channel to be a random variable $X$ and let $P(x)$ be the channel input probability mass function (pmf) and $P(y|x)$ the channel law, i.e., the
probability of receiving $Y = y$ when sending $X = x$. As we have stated in the introduction, we require that the distribution  $P(x)$ have this form
\begin{equation}
P(x)=\left\{
\begin{array}{ll}
1/K & \textrm {if }	x \in \mathcal{X}' \\
0 & \textrm {otherwise,}
\end{array}
\right.	
\end{equation}	
i.e., it is a uniform distribution over a subset $\mathcal{X}' \subset \mathcal{X}$. Different criteria can be considered to find the best subset $\mathcal{X}'$ for given size $|\mathcal{X}'|=K<M$:\\
a) Maximizing the mutual information $I(X,Y)$\\
\begin{equation}	
\max_{\mathcal{X}'\subseteq\mathcal{X}} \log_2 K+\frac{1}{K} \sum_y\sum\limits_{x\in \mathcal{X}'} P(y|x) \log_2\frac{P(y|x)}{\sum\limits_{x'\in \mathcal{X}'} P(y|x') }
\end{equation}	
b) Minimizing the symbol error rate (SER) assuming ML decoding
\begin{equation}	
\min_{\mathcal{X}'\subseteq\mathcal{X}}
 1 - \frac{1}{K}\sum_{y} \max_{x \in \mathcal{X}' } P(y|x)
\end{equation}
c) Maximizing the cut-off rate $R_0$
\begin{equation}	
 \max_{\mathcal{X}'\subseteq\mathcal{X}}´ 2\textrm{log}_2K-\textrm{log}_2\sum_y\left[\sum_{x \in \mathcal{X}'} \sqrt{P(y|x)}\right]^2.
 \label{c)}
\end{equation}
Note that in all these problems the subset size is assumed to be a priori fixed and higher than $2^{C}$, where $C$ denotes the true capacity.\footnote{Clearly the subset size have to be chosen properly; this aspect will be discussed later.}  Problem a) is the most interesting from the information theoretical point of view. However we were not able to find an efficient algorithm for solving this problem. 
Nevertheless, it turns out that all these criteria are very correlated, so that the solution of one problem  
is nearly-optimal for the others. Therefore, we consider in this work as alternative the SER minimization b) and the  cut-off rate maximization c) problems due to their more tractable structures. \\
Throughout our paper, $a_i$ denotes the $i$-th element of a vector $\boldsymbol{a}$ and $A_{ij}$ the element of a matrix $\B{A}$ in row $i$ and column $j$. The operators $(\cdot)^\textrm{T}$ and $\textrm{tr}(\cdot)$ stand for transpose and trace of a matrix, respectively. Vectors and matrices are denoted by lower and upper case italic bold letters. $\B{0}_M$  and $\B{1}_M$ stand  for the $M$-length zero and all ones vector respectively.
\section{MINIMIZING THE SYMBOL ERROR RATE}
\label{ser}
In this part, we look for the subset that minimizes the SER.
Note that this optimization task is an NP hard problem and an exhaustive search becomes intractable for large DMC. In \cite{zeger}, a binary switching algorithm (BSA), previously used for index optimization
in vector quantization, has been proposed
to overcome the complexity problems of the bruteforce
approach. This algorithm finds through systematic switch of symbols a local optimum on a given cost function. If the algorithm is executed several times with different random initializations, the global optimum may be found with high probability. The binary switching algorithm can be also used here to search for the optimal subset.
The input of the binary switching algorithm is the initial subset that is chosen randomly. The algorithm first generates  an ordered
list of the initially selected symbols, sorted according to the decreasing order of their costs  calculated for individual symbols (probability of misdetecting an $x$)
\begin{equation}	
 {\rm Pr}(x\rightarrow \hat{x}\neq x)=
  \sum_{y|x\neq\argmax\limits_{x' \in \mathcal{X'} } P(y|x')} \! \! \! \! P(y|x), \quad x \in \mathcal{X'}.
\vspace{-0.0cm}
\end{equation}
 Then the algorithm tries to replace the symbol that has the highest cost with another symbol from the remaining subset $\mathcal{X} \backslash \mathcal{X'}$, which is selected
such, that the decrease of the total cost due to the
switch is as large as possible. If no switch 
can be found for the symbol with the highest cost,
the symbol with the second-highest cost will be tried
to be replaced next. Also, if the lowest total cost is lower than the initial cost, the switching is selected and the iteration is continued, else we try the third one and so on. After
an accepted switch, a new ordered list of symbols is
generated, and the algorithm continues as described
above until no further reduction of the total cost is
possible. The BSA converges to a subset with a local optimal cost. To find the subset with global optimal cost, we can start the algorithm with different initializations several times and select the result with the lowest total cost.
\section{MAXIMIZING THE CUT-OFF RATE}
\label{section:cutoff}
The cutoff rate $R_0$ can be used for practical finite length block
codes in discrete memoryless channels to upper-bound codeword
error rates after maximum likelihood decoding. Besides, it represents a  lower bound on the channel capacity. Thus the maximization of the cutoff rate is essential to have good performance in practice. 
Problem (\ref{c)}) can be reformulated as
\begin{equation}
\begin{aligned}
& \min_{\B{b} \in \{0, 1\}^M}    \quad \B{b}^T\B{A}\B{b} \\
& \quad \textrm{ s.t. }  \B{1}^{\rm T}_M\B{b}=K,
\end{aligned}
\label{problem}
\end{equation}
where
\begin{equation}
\B{A}=\sum\limits_{y}\left[ 
\begin{array}{c}
	\sqrt{P(y|1)}\\
	\vdots \\
		\sqrt{P(y|M)}
\end{array}
\right] \cdot \left[ 
\begin{array}{c}
	\sqrt{P(y|1)}\\
	\vdots \\
		\sqrt{P(y|M)}
\end{array}
\right]^{\rm T},
\end{equation}
and the vector $\B{b}$ is a binary vector consisting only of the elements "0" and "1". The input symbols are here numbered consecutively from 1 to $M$. The ones in vector $\B{b}$ indicates the symbols included in the subset $\mathcal{X}'$. The formulation (\ref{problem}) is a constrained binary quadratic minimization problem (constrained BQP), thus we have to do with an NP-hard combinatoric problem. It can be interpreted as a two partitioning problem with fixed partition size. The matrix coefficient $A_{ij}$ can be interpreted as the cost of selecting the input $i$ and $j$ into the subset $\mathcal{X}'$.\\ 
Now, we introduce the vector $\B{s}=[s_1,\cdots,s_n]^{\rm T}$, with $n=M+1$, and relate it to $\B{b}$ as follows
\begin{equation}
\B{b}=s_n\cdot [s_1,\cdots,s_M]^{\rm T},
\end{equation} 
where the slack variable $s_n \in \{-1, 1\}$. This substitution is used to symmetrize the problem, which is necessary for the later convex problem formulation.\footnote{If $\B{s}$ is optimal then also is $-\B{s}$.} Then, it can be shown that problem (\ref{problem}) is equivalent to
\begin{equation}
\begin{aligned}
&\min_{\B{s}}    \quad \B{s}^T\B{B}\B{s} \quad \textrm{s.t. } \\
  s_n \cdot \B{1}^{\rm T}_n  \B{s}=& K+1 \textrm{, } s_n^2=1\textrm{, } s_{i}s_{n}-s_{i}^2=0 ~ \forall i,
 \end{aligned}
\label{problem1}
\end{equation}
with
\begin{equation}
\B{B}=\left[
\begin{array}{cc}
	\B{A} &  \B{0}_M \\
	\B{0}_M^{\rm T } & 0
\end{array}\right].
\end{equation}
 By means of the substitution $\B{S}=\B{s}\B{s}^{\rm T}$, where $\B{S}$ is a positive semidefinite matrix  ($\B{S} \succeq \B{0}$) of rank 1, problem (\ref{problem1})  can be rewritten into the matrix optimization problem 
\begin{equation}
\begin{aligned}
\hat{\B{S}}&=\argmin\limits_{\B{S} \succeq \B{0}} {\rm tr}(\B{B} \B{S})   \textrm{ s.t.}\\
S_{ii}=S_{in}& ~ \forall i, S_{nn}=1 ,  \sum\limits_{i}  S_{ni}=K+1  \\
 &\textrm{ and }    {\rm rank}(\B{S})=1.
 \end{aligned}
\label{problem2}
\end{equation}
The program (\ref{problem2}) is not convex because of the rank-one constraint. Recently, semidefinite
programming (SDP) has been shown to be a very promising
approach to combinatorial optimization, where SDP serves as a
tractable convex relaxation of NP-hard problems. 
 In \cite{steingrimsson}, for example, a quasi-maximum likelihood method based on Semi-Definite Programming (SDP) for lattice decoding is introduced. \\
 In order to obtain a tractable SDP relaxation of (\ref{problem2}), we remove the rank-one
restriction from the feasible set
\begin{equation}
\begin{aligned}
\hat{\B{S}}&=\argmin\limits_{\B{S}  \succeq \B{0}} {\rm tr}(\B{B} \B{S})   \textrm{ s.t.}\\
S_{ii}=S_{in}&~ \forall i, S_{nn}=1 ,  \sum\limits_{i}  S_{ni}=K+1.  \\
 \end{aligned}
\label{problem3}
\end{equation}
Note that this optimization  has a linear objective subject to affine equalities
and a linear matrix inequality. Such problems are known as
SDP and can be efficiently solved in polynomial time \cite{sturm}.\footnote{It is possible to  solve SDP relaxations of boolean QPs for problems of fairly large
size (approx. 500 vars with interior point, 5000+ with special techniques).} \\
If the optimal solution  of the  SDP has rank one, then the
relaxation is tight. Otherwise, some special techniques are required to convert the SD relaxation solution to an approximate Boolean QP solution. A randomization method has been proposed for this conversion process \cite{nesterov}.  This is motivated via a probabilistic argument.
For this, assume that rather than choosing the optimal $\B{s}$ in a deterministic fashion, we want to find  instead a probability distribution with covariance matrix $\B{S} ={\rm E}[\B{s}^{\rm T}\B{s}]$ that will yield good solutions on average. For symmetry reasons, we can always restrict ourselves to distributions with zero mean. For the
constraints, we may require that the solutions we generate fulfill the constraints on expectation. Maximizing the expected value of the cost (\ref{problem1}), under average constraints  yields the SDP relaxation presented in (\ref{problem3}). 
\begin{algorithm}
\caption{Codebook Selection Algorithm}
\label{Codebook Selection Algorithm}
\begin{algorithmic}[1]
\STATE \textbf{Initialization:} $n=M+1$ \\
\vspace{0.2cm}
$\B{A}=\sum\limits_{y}\left[ 
\begin{array}{c}
	\sqrt{P(y|1)}\\
	\vdots \\
		\sqrt{P(y|M)}
\end{array}
\right] \cdot \left[ 
\begin{array}{c}
	\sqrt{P(y|1)}\\
	\vdots \\
		\sqrt{P(y|M)}
\end{array}
\right]^{\rm T}$\\
\vspace{0.2cm}
$\B{B}=\left[
\begin{array}{cc}
	\B{A} & \B{0}_M \\
	\B{0}_M^{\rm T } & 0
\end{array}\right]
$
\vspace{0.2cm}
\STATE \textbf{Solve the semi-definite problem:} \\
\begin{center}
  $\hat{\B{S}}=\argmin\limits_{\B{S} \succeq \B{0}} {\rm tr}(\B{B} \B{S})$   s.t.\\
 $S_{nn}=1$, $\sum\limits_{i} S_{ni}=K+1$ and $S_{ii}=S_{in} ~ \forall i$
  \end{center}
\STATE  \textbf{Cholesky factorization:} $\hat{\B{S}}=\hat{\B{V}}^{\rm T}\hat{\B{V}}$
\STATE \textbf{Randomization:} 
\FOR{$i$ = $1,\ldots,N_{\rm rand}$}
\STATE Randomly generate a vector $\B{u}^{(i)}$ uniformly distributed on a $n$-dimensional unit sphere. 
\STATE Compute $\tilde{\B {s}}^{(i)}=\hat{\B{V}}^{\rm T} \B{u}^{(i)}$ $ ~ \forall i$. 
\STATE $\tilde{s}_n^{(i)}  \leftarrow  {\rm sign} (\tilde{s}_n^{(i)})$
\STATE $\tilde{\B {s}}^{(i)} \leftarrow \tilde{s}_n^{(i)}  \tilde{\B {s}}^{(i)}$ 
\STATE Quantize the $K$ highest entries of $[\tilde{s}_1^{(i)}, \cdots, \tilde{s}_{n-1}^{(i)} ]$ to 1 and the others to 0
\ENDFOR  
\STATE  \textbf{Choose} $ \tilde{\B {s}}= \argmax\limits_{\tilde{\B {s}}^{(i)}} \tilde{\B {s}}^{(i),{\rm T}}  \B{B}  \tilde{\B {s}}^{(i)}$
\STATE   \textbf{Take} $\B{b}=[\tilde{s}_0, \cdots, \tilde{s}_{n-1}]^{\rm T}$ as approximate solution
\end{algorithmic}
\end{algorithm}
\par Usually, to further improve the approximation quality, the randomization is repeated a number of 
times, and the randomized solution yielding the largest objective function value is chosen as the 
approximate solution. This procedure is stated in Step 4 to 12 of Algorithm \ref{Codebook Selection Algorithm}. Often, this randomization 
method can achieve an accurate approximation with a modest number of randomizations.
An other more simple approach consists in taking $\tilde{\B{s}}$ as  the eigenvector
of $\hat{\B{S}}$ associated with its maximal eigenvalue and then simply performing the quantization procedure (step 8, 9, 10), which can also provide good solutions.
\section{APPLICATION TO COARSELY QUANTIZED MIMO}
\label{application}
\begin{figure}[h]
\begin{center}
\psfrag{H}[c][c]{$\boldsymbol{H}$}
\psfrag{G}[c][c]{$\boldsymbol{G}$}
\psfrag{xd}[c][c]{$\boldsymbol{\hat{x}}$}
\psfrag{x}[c][c]{$\boldsymbol{x}$}
\psfrag{y}[c][c]{$\boldsymbol{y}$}
\psfrag{r}[c][c]{$\boldsymbol{r}$}
\psfrag{Q[.]}[c][c]{$Q(\bullet)$}
\psfrag{n}[c][c]{$\boldsymbol{\eta}$}
\psfrag{M}[c][c]{$T$}
\psfrag{N}[c][c]{$N$}
\psfrag{SNR}[c][c]{$\sqrt{P_\textrm{Tr}}$}
{\epsfig {file=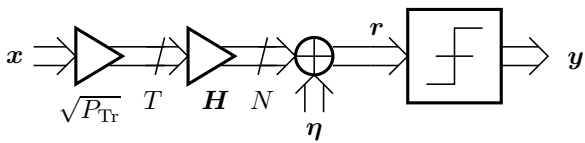, width = 7.5cm}}
\caption{One-bit Quantized MIMO System}
\label{downlink_figure}
\end{center}
\end{figure}

As application, we consider a  point to point mono-bit quantized MIMO system for high speed links \cite{nossek}, where the transmitter employs $T$ antennas and the  receiver has $N$ antennas. Fig.~\ref{downlink_figure} shows the general form of a quantized MIMO system, where $\boldsymbol{H} \in \mathbb{C}^{N\times T}$ is the channel matrix, known at both the transmitter and the receiver. We assume each entry $x_i$ of the source symbol $\B{x}$ is drawn from a discrete QPSK modulation, so that the source alphabet $\mathcal{X}$ has a cardinality of $|\mathcal{X}|=M=4^T$. The average energy of $\boldsymbol{x}$ is fixed to 1, i.e., $\B{x}^{\rm H} \B{x}=1$. The vector $\boldsymbol{\eta}$ refers to uncorrelated zero-mean complex circular Gaussian noise with equal variance per dimension given by $\sigma_\eta^2$. The unquantized channel output $\boldsymbol{r}\in\mathbb{C}^{N}$ is given by
\begin{equation}
 \boldsymbol{r}=\sqrt{P_\textrm{Tr}}\boldsymbol{H}\boldsymbol{x}+\boldsymbol{\eta},
\end{equation}
where $P_\textrm{Tr}$ is the transmit power.\\
 In this system, the real parts $r_{i,R}$ and the imaginary parts $r_{i,I}$ of the receive signals $r_i$, $1\leq i\leq N$, are each quantized by a $1$-bit resolution quantizer. Thus, the resulting quantized signals read as:
\begin{equation}
y_{i,c}=\textrm{sign}(r_{i,c})\in\{-1,1\},\textrm{ for }c\in\{R,I\},\textrm{ }1\leq i\leq N.
\end{equation}
Obviously the scalar (complex) quantization of the output of the QPSK MIMO channel with hard decision receivers produces an equivalent channel with $4^T$ inputs and $4^{N}$ outputs. The resulting channel can be seen as  a large strongly non-symmetric Discrete Memoryless Channel (DMC)  \cite{nossek}, and characterized by a transition probability matrix   $P(\B{x}|\B{y}) $.
Since all of the real and imaginary components of the receiver noise $\boldsymbol{\eta}$ are statistically independent with variance ${\sigma_\eta^2}/{2}$, we can express each of the conditional probabilities as the product of the conditional probabilities on each receiver dimension
\begin{equation}
\begin{aligned}
P(\boldsymbol{y}|\boldsymbol{x})&=\prod_{c\in\{R,I\}}\prod_{i=1}^{N}P(y_{c,i}|\boldsymbol{x})\\
&=\prod_{c\in\{R,I\}}\prod_{i=1}^{N}\Phi\left(\sqrt{2P_\textrm{Tr}/\sigma_\eta^2}y_{c,i}[\boldsymbol{H}\boldsymbol{x}]_{c,i}\right),
\label{cond_pro}
\end{aligned}
\end{equation}
with $\Phi(x)=\frac{1}{\sqrt{2\pi}}\int_{-\infty}^{x}e^{-\frac{t^2}{2}}dt$ is the cumulative normal distribution function.\\
As example, we
use a random generated MIMO channel matrix specified as:
\begin{equation}
{\rm Re}[\B{H}] =  \frac{1}{100}
\left[
\begin{array}{r r r r}
-31 &-81& 23& -42 \\
-118 &84 &12 &154  \\
9 &84& -13& -10   \\
20 &-3 &51 &8  
\end{array} \right],
\end{equation}
\begin{equation}
{\rm Im}[\B{H}]= \frac{1}{100}
\left[
\begin{array}{r r r r}
75 & 21 & -49 & -102  \\
4 & -94 & 61 & 40   \\
-7 & 51 & 89  & -28   \\
-59 & 115 & -113 & 49 
\end{array}
 \right]  .
\end{equation}
\begin{figure}[h]
\begin{center}
\psfrag {Capacity}[l][l]{Capacity}
\psfrag {SNR}[l][l]{$P_{\rm Tr}/\sigma_\eta^2$ (dB)}
\psfrag {True Capacity}[l][l]{True Capacity}
\psfrag{Uniform Capacity}[l][l]{Uniform Capacity}
\psfrag{16S Uniform Capacity}[l][l]{$K=16$, Capacity}
\psfrag{64S Uniform Capacity}[l][l]{$K=64$, Capacity}
{\epsfig{file=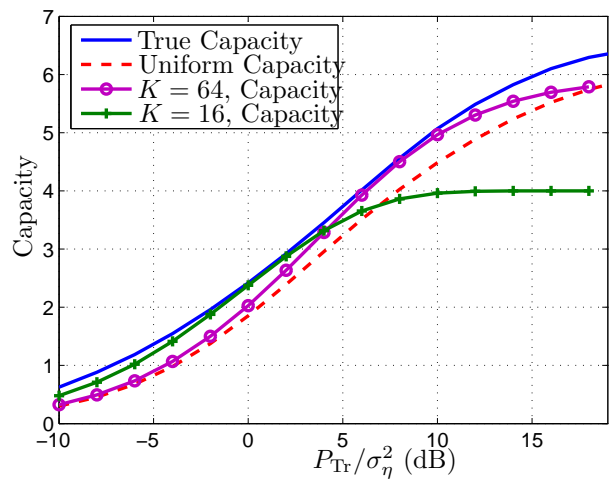, width = 9.4cm}}
\caption{Capacity improvement of 4x4 QPSK MIMO with mono-bit receiver under $K=16$ and $K=64$ selected symbols found by algorithm 1.}
\label{transinfo_fig}
\end{center}
\end{figure}
The solid line in Fig.~\ref{transinfo_fig} shows the true capacity of this channel obtained by optimizing the input distribution using the Blahut-Arimoto algorithm \cite{cover}. The capacity achieved by the uniform prior over all symbols is also plotted (dashed line), where a considerable rate loss can be observed. Now, applying Algorithm \ref{Codebook Selection Algorithm} to this channel for two different values of $K$ ($K=64$ and $K=16$) leads to the marked solid curves. The semidefinite program in the SDP was solved using
the SeDuMi package \cite {sturm}.
\begin{figure}[h]
\begin{center}
\psfrag {Capacity}[l][l]{Capacity}
\psfrag {SNR}[l][l]{$P_{\rm Tr}/\sigma_\eta^2$ (dB)}
\psfrag {True Capacity}[l][l]{True Capacity}
\psfrag{Uniform Capacity}[l][l]{Uniform Capacity}
\psfrag{16S Uniform Capacity}[l][l]{$K=16$, Capacity}
\psfrag{64S Uniform Capacity}[l][l]{$K=64$, Capacity}
{\epsfig{file=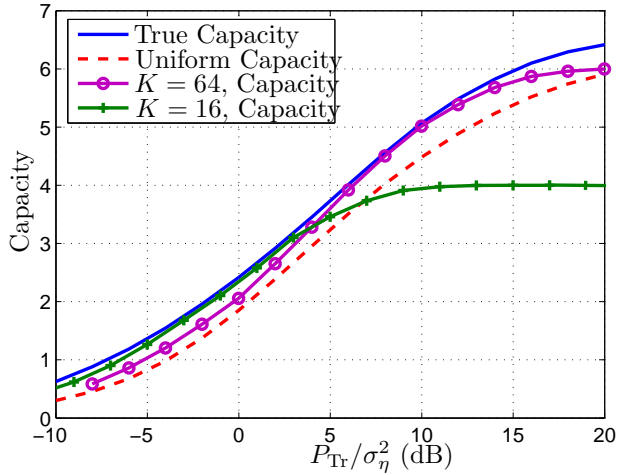, width = 9.4cm}}
\caption{Capacity improvement of 4x4 QPSK MIMO with mono-bit receiver under $K=16$ and $K=64$ selected symbols using the SER minimization solved by the BSA.}
\label{transinfo_fig2}
\end{center}
\end{figure}
 Although the selection  is based on the  cut-off criterion, the resulting subsets almost achieve  the capacity on a quite large SNR interval. It seems that the two subset sizes are sufficient to cover a wide dynamic range of the SNR. Besides it turns out that the optimal subset doesn't depend strongly on the SNR. This shows the usefulness of this approach. \\
 Figure~\ref{transinfo_fig2} shows the capacity results obtained for the codebooks selected based on the SER criterion and the BSA as described in Section~\ref{ser} under the same settings. Obviously the results are very similar to those in Fig~\ref{transinfo_fig2}, which confirms our previous  hypothesis that the selection does not depend strongly on the chosen criteria. We note that the convergence time of the BSA depends on the channel
conditions and the noise level and it may become useless for larger DMC. All in all, it is preferable to employ algorithm~1 rather than the BSA, since its convergence time is fixed and only polynomial in the size of $\mathcal{X}$. 
\vspace{0cm}
\section{PERFORMANCE WITH CODING} 
 \label{ldpc}
\begin{figure}[h]
\begin{center}
\psfrag {Capacity}[l][l]{Capacity}
\psfrag {SNR}[l][l]{$\!\!\!P_{\rm Tr}/\sigma_\eta^2$ (dB)}
\psfrag{BER}[l][l]{Coded BER}
\psfrag{64S Uniform Capacity}[l][l]{$K=64$, Capacity}
{\epsfig{file=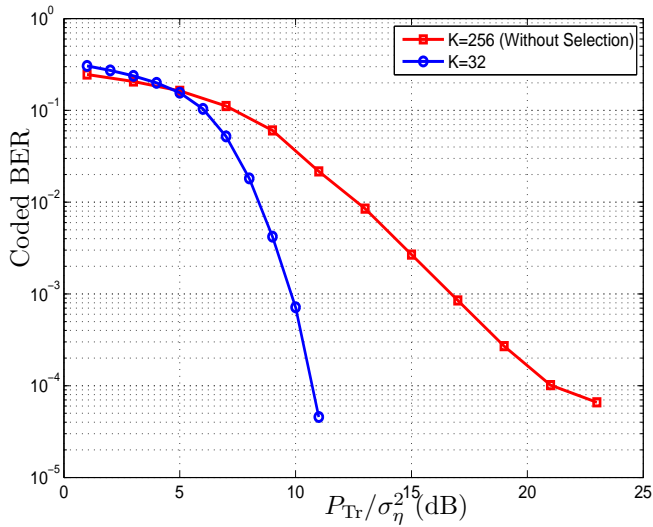, width = 9cm, height=7cm }}
\caption{Bit error ratio of LDPC code after decoupled log-likelihood computation and belief-propagation algorithm.}
\label{ber}
\end{center}
\end{figure}

Approaching the channel capacity of coarsely quantized
MIMO systems is however not straight forward. Figure~\ref{ber}
shows the bit error ratio obtained with an ensemble of randomly generated LDPC code of length $n=250$ applied on  the same channel as in the previous section. The parity check matrices  were generated following \cite{mackay}.
The performance of our input set reduction method with $K=32$ compared to the full input use ($K=256$) in terms of BER when combined with an LDPC code is shown in this figure. For both cases the total rate is $R=2.5$ bits/channel use; and the rate of the LDPC code was adjusted for each case accordingly. 
 We apply a decoupled detection/decoding approach, where first the
log-likelihood ratios
\begin{equation}
\log \left( \frac{{\rm Pr}[ c[i] = 1 | y[n] ]}
{1 - {\rm Pr}[ c[i] = 1 | y[n] ]}\right)
\end{equation}
are computed and then fed to the input of the belief-propagation algorithm.
Here $c[i]$ denotes the $i$-th bit that is output by
the LDPC encoder, while $\B{y}[n]$ is the $n$-th quantized
received vector, where
\begin{equation} 
n = {\rm floor}(i/\log_2 K).
\end{equation}
This comes about, since $\log_2 K$ code-bits are transfered per channel
use, hence, for each received quantized vector $\B{y}[n]$ the
log-likelihood ratios of $\log_2 K$ encoded bits are computed.
Obviously the proper reduction of the input set improves the BER behavior significantly. Besides the full use of the input set 
  cannot be handled
gracefully, leading to a relatively large error floor. This is caused by the fact that with coarse channel
 output quantization, many different input symbols may be assigned to the same output symbols at high SNR. 
  To resolve this ambiguities small code rate and large block length would be necessary, which leads again to high latency time and complex receiver. Fortunately, reducing the input set solves  this problem in a simpler way. As we see in Fig.~\ref{ber}, the optimal constellation does not see any error floor and the receiver's task become  easier with the more distinguishable selected symbols.
\section{CONCLUSION}
\label{section:conclusion}
A method is proposed that allows approaching the true capacity of large DMC channels while using uniformly distributed reduced input set. This has essential practical aspects since it allows the use of binary codes to approach the capacity without distribution shapers.
In addition, the idea of reducing the input to symbols that are maximally spaced makes the task of the decoder considerably easier and  inherently includes some robustness against the quality of the channel state information at the transmitter and other parameter fluctuation (SNR) in the system. To find the optimal input subset, we explored among others 
 SDP relaxation techniques, that turns to be a very efficient approach providing excellent solutions for this problem.
\bibliographystyle{IEEEbib}
\setlength{\textheight}{15 cm}
\bibliography{IEEEabrv,references}
\end{document}